\documentclass[12pt]{article}

\usepackage{amsmath,amssymb} %>>> for AMS math formatting, 
\usepackage[]{graphicx}
\usepackage{epstopdf}
\usepackage{multirow}
\usepackage{color}
\usepackage{lineno}
\usepackage{ulem}
\usepackage{upgreek}
\usepackage{lipsum}
 
\advance \textwidth by 1.1cm
\advance \textheight by 0.66cm
\def\eps{\varepsilon}
\newcommand{\fraz}{\displaystyle\frac}
\def\##1{{\bf #1}}
\def\=#1{\underline{\underline #1}}
\def\ux{\#u_x}
\def\uy{\#u_y}
\def\uz{\#u_z}

\def\muo{\mu_{\scriptscriptstyle 0}}
\def\epso{\eps_{\scriptscriptstyle 0}}
\def\lambdao{\lambda_{\scriptscriptstyle 0}}
\def\etao{\eta_{\scriptscriptstyle 0}}
\def\ko{k_{\scriptscriptstyle 0}}
\def\fo{\nu_{\rm c}}
\def\Bo{B_{\scriptscriptstyle 0}}
\def\Bov{\#B_{\scriptscriptstyle 0}}
\newcommand{\SImum}{\ensuremath{\upmu}\textrm{m}}

\def\tond#1{\left(#1\right)}
\def\quadr#1{\left[#1\right]}
\def\graff#1{\left\{#1\right\}}
\def\mod#1{\left|#1\right|}

\def\epsa{\eps_\perp}
\def\epsb{\eps_\parallel}
\def\epsab{\eps_\times}

\begin{document}
\begin{center}
{\bf Bicontrollable Terahertz Metasurface with Subwavelength Scattering Elements of 
Two Different Materials}\\

Francesco Chiadini$^1$ and
 Akhlesh Lakhtakia$^2$ \\

$^1$ Department of Industrial Engineering,
	University of Salerno, via Giovanni Paolo II, 132-Fisciano (SA), 
	I-84084, Italy\\
$^2$ Department of Engineering Science and Mechanics, Pennsylvania State 
University,
	University Park, PA 16802--6812,
	USA
	
	\end{center}

\begin{abstract}

Transmission of a normally incident plane wave through a metasurface
with bicontrollable subwavelength scattering elements was simulated using a
commercial software. Some pixels comprising the $\sf{H}$-shaped scattering elements were made of  a magnetostatically controllable material whereas
the remaining pixels were made of a thermally controllable material, the metasurface designed to operate in the terahertz spectral regime.
The co-polarized transmission coefficients were found
to exhibit stopbands that shift when either a magnetostatic field is applied 
or the temperature is increased or both.  Depending on spectral location of
the stopband, either
the magnetostatic field gives coarse control and temperature gives fine 
control or \textit{vice versa}. The level of magnetostatic control depends on the
magnetostatic-field configuration.
\end{abstract}

\section{Introduction} \label{sec:intro}

A frequency-selective-surface (FSS) is a planar periodic array of identical  scattering elements \cite{Wu,Munk}. The elements are electrically thin in the direction normal
to the surface. Each element is either infinitely long in one direction on the surface  or has finite dimensions in any direction on the surface. Typically, FSSs are used
as  bandpass and bandstop filters as well as to redirect a plane wave in a nonspecular
direction.

If the lattice parameters (and, therefore, the linear dimensions of the
scattering elements) of a 2D FSS are  sufficiently small fractions of the
free-space wavelength $\lambdao$, and the wave vector of the incident plane wave
is directed not very obliquely,  the transmitted field
contains a  plane wave propagating in the specular direction along with a
multitude of  evanescent plane waves that are nonspecular. The 2D FSS is then more commonly referred to as a metasurface these days \cite{Holloway,Veysi,Chen,Serebryan}.

Metasurfaces are useful in bandstop filters \cite{Han,Min},  absorbers \cite{Padilla},
and polarimeters~\cite{Capas}. Metasurfaces that generate local phase changes can be used to control the wavefront \cite{Jia,Hua}. Application to holography
is another area of research interest \cite{Hack}. Thus, metasurfaces are attractive for  molding  electromagnetic-wave propagation in a variety of ways.

This attraction is quite pronounced in the terahertz (THz) regime. Within this spectral regime, significant benefits
can be realized for power electronics \cite{Sinha}, imaging \cite{Suen}, spectroscopy \cite{Qin},   sensing \cite{Bhatt},
cancer detection \cite{cancer}, and so on. THz imaging
has immense potential for security technology, as it can extract the spectroscopic fingerprints of a wide range
of chemicals used in explosives and biological weapons \cite{Miles}.

The concept of multicontrollable metasurfaces has recently been put forth, with inspiration from  biological examples of multicontrollability \cite{AkhSPIE}. Each subwavelength scattering element in a multicontrollable metasurface is  a set of non-overlapping pixels. All pixels in a specific subset are made of a specific material, and the scattering element is made of at least two different materials. The electromagnetic constitutive parameters of each of these materials in the chosen spectral regime
 may be controlled by the variation of a specific environmental parameter such as   temperature,   voltage, and a magnetostatic field. Thus, the overall electromagnetic response characteristics of a multicontrollable metasurface can be dynamically controlled using one or more  modalities.

In order to establish the feasibility of multicontrollable metasurfaces, in this paper we present theoretical results on the THz transmission characteristics of a bicontrollable metasurface whose scattering elements comprise  magnetostatically controllable pixels made of InAs \cite{Han,Serebryan} and thermally controllable pixels made of CdTe \cite{Schall}. The chosen metasurface and the simulation method are described in Sec.~\ref{sec:matmeth}  
Numerical results of transmission simulations  are presented in
Sec.~\ref{sec:RaD} to establish  the desired bicontrollability. Conclusions follow under Sec.~\ref{sec:cr}.

An $\exp\tond{-i\omega t}$  dependence on time $t$ is implicit, with $\omega=2\pi\nu$ denoting the angular frequency, $\nu$ the linear
frequency, and $i=\sqrt{-1}$. The free-space wavenumber and the intrinsic impedance of free space are denoted by $\ko=\omega\sqrt{\epso \muo}=2\pi/\lambdao$ and $\etao=\sqrt{\muo/\epso}$, respectively, with $\epso$ and $\muo$ being the permittivity and permeability   of free space. Vectors are in boldface; dyadics are underlined twice; and  the three Cartesian unit vectors are identified as $\ux$, $\uy$, and $\uz$.

\section{Materials and Method}\label{sec:matmeth} 

\begin{figure}[htbp]
	\begin{center}
		\begin{tabular}{c}
			(a)\includegraphics[width=8cm]{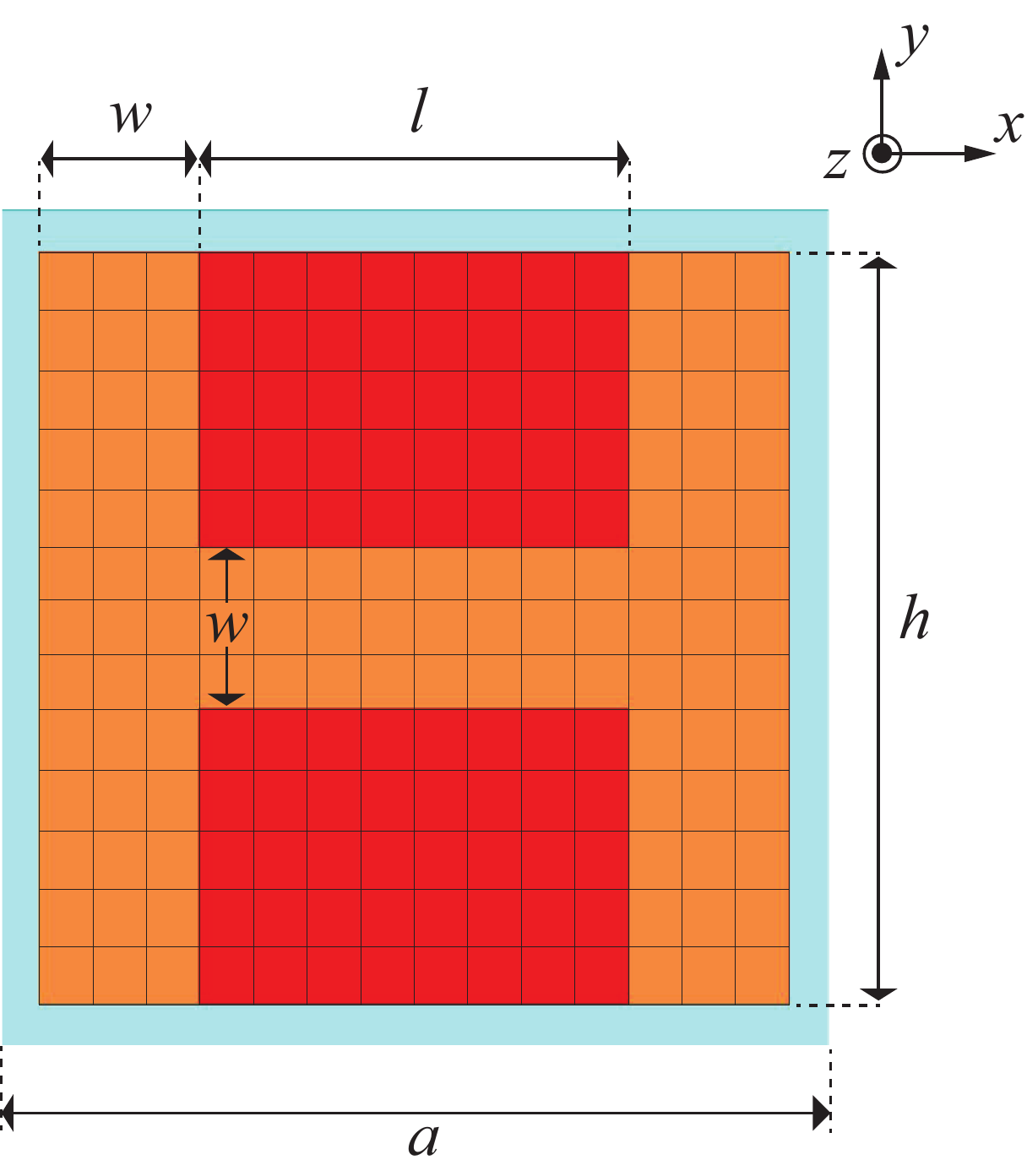}\\ %\hspace{20pt}
			(b)\includegraphics[width=8cm]{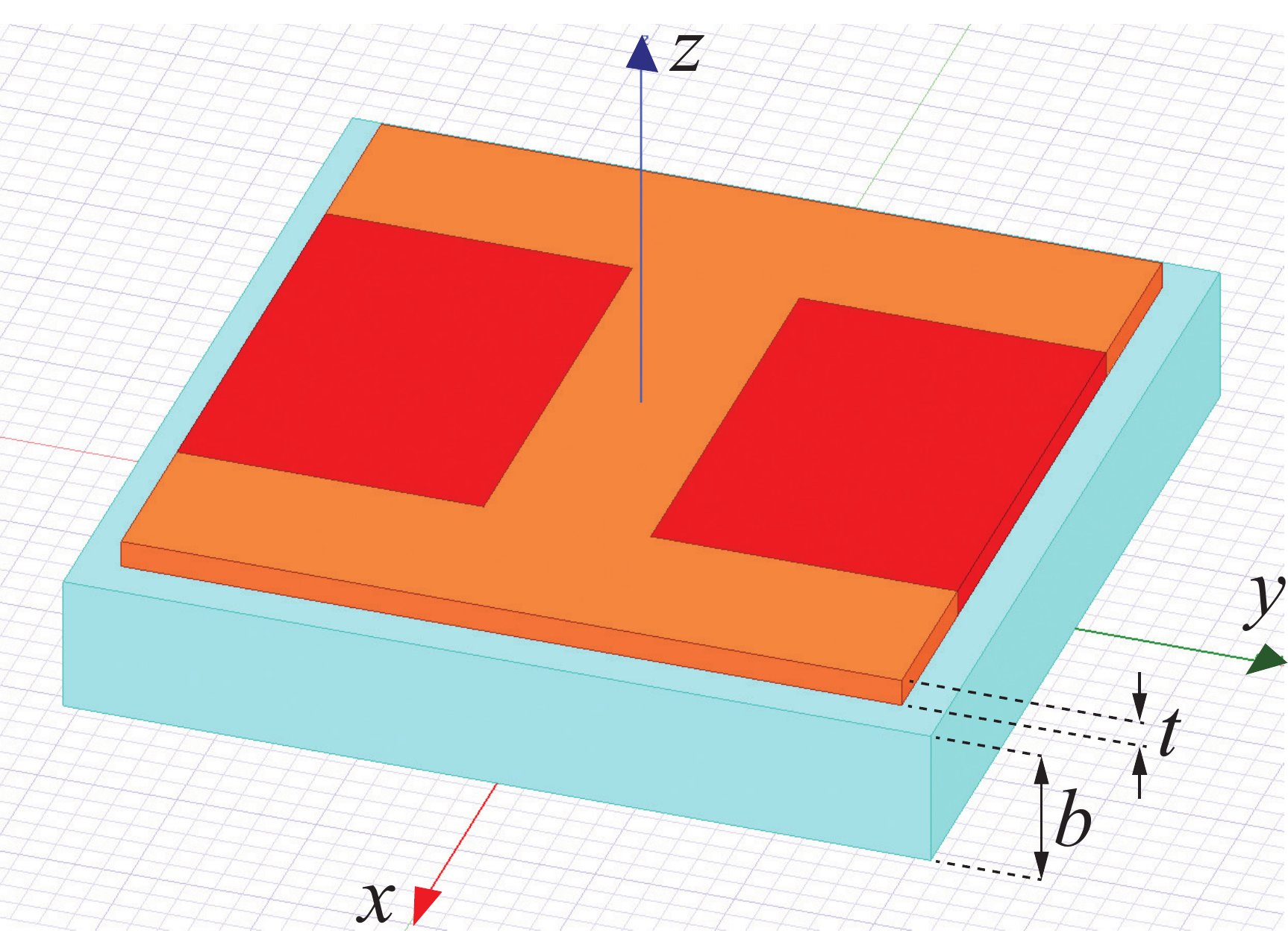}
		\end{tabular}
	\end{center}
	\caption{(Color online) (a) Top and (b) isometric views of the unit cell of the chosen bicontrollable metasurface comprising magnetostatically controllable (orange) and thermally controllable (red) pixels deposited on an inert substrate (blue). Various linear dimensions as well as the Cartesian axes are also shown.}
	\label{fig:schem}
\end{figure}

The unit cell of the chosen bicontrollable metasurface is
a rectangular parallelepiped of dimensions $a\times{a}\times(t+b)$
aligned with the $x$, $y$, and $z$ axes,
as shown in Fig.~\ref{fig:schem}. The substrate occupies the region $z\in(-t-b,-t)$
and is made of an isotropic material with relative permittivity $\eps_d$. The active region of the unit cell is
a rectangular parallelepiped   of dimensions $(2w+l)\times{h}\times{t}$
aligned with the $x$, $y$, and $z$ axes, with $(2w+l)\leq{a}$, $h\leq{a}$, and
$t\ll{a}$. The active region is made of 
\begin{itemize}
\item[(i)] two $w\times{h}\times{t}$ sections and one
$l\times{w}\times{t}$ section, together forming an $\sf{H}$
and all three comprising pixels made of
a magnetostatically controllable material, and
\item[(ii)] two $l\times{(1/2)(h-w)}\times{t}$ sections   comprising pixels made of
a thermally controllable material.
\end{itemize}
For calculations, we chose  InAs as 
the magnetostatically controllable material~\cite{Serebryan,Han}
and  CdTe as the thermally controllable material~\cite{Schall}.

%%%%%%%%%%%%%% FIGURE 1 %%%%%%%%%%%%%%%%%%%%%%%%
%%%%%%%%%%%%%%%%%%%%%%%%%%%%%%%%%%%%%%%%%%%%%%%%

InAs is an isotropic dielectric material in the absence of an external 
magnetostatic field and its relative permittivity can be appropriately represented
by the
Drude model \cite{Blaber}. When it is subjected to a magnetostatic field $\Bov$, InAs functions as a
 gyroelectric material~\cite{Han,Serebryan}.  Its relative
 permittivity dyadic then depends on the magnitude $\Bo=\vert\Bov\vert$
 and the direction $\Bov/\Bo$
 of  $\Bov$.
 The following three configurations are distinctive: 
 \begin{itemize}
 \item[(i)] Faraday configuration ($\Bov\parallel\uz$), with the relative
 permittivity dyadic
\begin{equation}
\={\eps}_{InAs}^F=
\epsa\tond{\ux\ux+\uy\uy} +\epsb\uz\uz+ \epsab\tond{\ux\uy -\uy\ux} \,;
\label{eq:eps_InAs_F}
\end{equation}
 \item[(ii)] Voigt-X configuration ($\Bov\parallel\ux$), with the relative
 permittivity dyadic
\begin{equation}
\={\eps}_{InAs}^{VX}=\epsa\tond{\uy\uy+\uz\uz}+
\epsb\ux\ux+ \epsab\tond{\uy\uz-\uz\uy}
\,;
\label{eq:eps_InAs_Vx}
\end{equation}
and
 \item[(iii)] Voigt-Y configuration ($\Bov\parallel\uy$), with the relative
 permittivity dyadic
\begin{equation}
\={\eps}_{InAs}^{VY}=
\epsa\tond{\ux\ux+\uz\uz} +\epsb\uy\uy+\epsab\tond{\uz\ux-\ux\uz} \,.
\label{eq:eps_InAs_Vy}
\end{equation}
\end{itemize}
In
\begin{equation}
\left.\begin{array}{l}
\epsa=\eps_{\infty}-\fraz{\bar{\omega}_p^2\tond{1+i\bar{\gamma}}}{\tond{1+i\bar{\gamma}}^2-\bar{\omega}_c^2}
\\[8pt]
\epsb=\eps_{\infty}-\fraz{\bar{\omega}_p^2}{\tond{1+i\bar{\gamma}}}\,,
\\[8pt]
\epsab=i\fraz{\bar{\omega_c}\bar{\omega}_p^2}{\tond{1+i\bar{\gamma}}^2-\bar{\omega}_c^2}
\end{array}
\right\}\,,
\end{equation}
$\eps_{\infty}=16.3$ is the value assumed by the relative 
permittivity 
in the limit $\omega\rightarrow\infty$; $\bar{\omega}_p=\omega_p/\omega$ is the 
normalized plasma frequency with $\omega_p=\sqrt{N_e q_e^2/\epso m^\ast}$ 
as the plasma frequency, $N_e=1.04\times 10^{23}$~m$^{-3}$ as the free-career 
density, $m^\ast=4\times 10^{-3} m_e$ as the effective carrier mass, 
$m_e=9.11\times 10^{-31}$~kg as the electron mass, and 
$q_e=1.6\times 10^{-19}$~C 
as 
the elementary charge; $\bar{\gamma}=\gamma/\omega$ is the normalized damping 
constant with $\gamma=15\pi\times 10^{11}$~rad~s$^{-1}$ as the damping constant; 
$\bar{\omega}_c=\omega_c/\omega$ is the normalized cyclotron frequency 
with $\omega_c=q_e \Bo/m^\ast$ as the cyclotron frequency that provides the 
dependence of the relative permittivity dyadic on the magnitude $\Bo$ of the magnetostatic field.
In this paper, we  provide  results for $\Bo=0$~T and $\Bo=1$~T to simulate an  on-off 
switching scenario.

CdTe is an isotropic dielectric material whose relative permittivity changes as a 
function of the temperature in the THz  regime. However, CdTe is  also an
electro-optic material so that we invoked the Pockels effect by the application of an electrostatic field ~\cite{Valagiann,Yariv}. When CdTe is subjected
to a dc electric field $\#{E}_{dc}=E_{dc}\uz$, its acts like an
orthorhombic dielectric material with relative permittivity dyadic
\begin{equation}
\={\eps}_{CdTe}=
\eps_\alpha\tond{\ux\ux+\uy\uy} +\eps_\beta\tond{\ux\uy+\uy\ux} +\eps_{\gamma}\uz\uz
\label{eq:eps_CdTe}
\end{equation}
where
\begin{equation}
\left.\begin{array}{l}
\eps_\alpha=\fraz{\eps_{\gamma}}{1-r_{\gamma}^2E_{dc}^2 \eps_{\gamma}^2}
\\[8pt]
\eps_\beta=-\fraz{r_{\gamma} E_{dc} \eps_{\gamma}^2}{1-r_{\gamma}^2E_{dc}^2 \eps_{\gamma}^2}
\end{array}
\right\}\,,
\end{equation}
 $r_{\gamma}=6.8\times 10^{-12}$~m~V$^{-1}$ is the  electro-optic coefficient,
and $\eps_{\gamma}$ is the relative permittivity in the absence of dc electric
field. Due to temperature dependence, we have \cite{Schall}
\begin{equation}
\eps_{\gamma}=\eps_{hf} 
+\fraz{\tond{\eps_{dc}-\eps_{hf}}\omega_T^2}{\omega_T^2 
-\omega^2-i\gamma_T\omega}
\label{eq:eps_CdTe_T}
\end{equation}
where $\eps_{hf}=6.8$ is the high-frequency relative permittivity,  $\omega_T$ is 
the resonance angular frequency,  
$\gamma_T$ is the damping constant, and $\eps_{dc}$ is the static relative 
permittivity. The latter three parameters can be 
expressed as  quadratic functions of the absolute temperature $T$ as 
\begin{equation}
Y =A+B\tond{\fraz{T}{T_{ph}}}+C\tond{\fraz{T}{T_{ph}}}^2\,,
\qquad Y\in\left\{\omega_T,\gamma_T,\eps_{dc}\right\}\,,
\label{eq:empir}
\end{equation}
where $T_{ph}=207$~K is the characteristic phonon temperature of CdTe.
The coefficients $A$, $B$, and $C$ are given in  Table \ref{tab:ABC}
for the application of Eq.~(\ref{eq:empir}) to $\omega_T$, $\gamma_T$,
and $\eps_{dc}$. In order to calculate all
numerical results presented here, we set  $E_{dc}=10^6$~V~m$^{-1}$ which is far below 
the  dielectric-breakdown limit $10^7$~V~m$^{-1}$ of CdTe. 

  %%%%%%%%%%%%%% TABLE 1 %%%%%%%%%%%%%%%%%%%%%%%%
\begin{table}
	\caption{\bf Coefficients of quadratic thermal dependences in the 
	Eq.~(\ref{eq:empir}) for $\omega_T$, $\gamma_T$, and 
	$\eps_{dc}$ of CdTe.\label{tab:ABC}
	}
	\begin{center} 
		\begin{tabular}{c|rrrr}
		Parameter	& $A$ & $B$ & $C$ &     Units\\ \hline
	$\omega_T\times10^{-12}$  & $27.401$ & $-0.1872$ & $-0.2187$ & rad~s$^{-1}$  \\
			 {$\gamma_T/\omega_T$} &  {$0.0116$} &   {$0.0314$}  & {$0.0119$} & \\
			$\eps_{dc}$ & $9.808$ &  {$0.1719$} & $0.1414$ &\\
			\hline%\hline
		\end{tabular}
	\end{center}
\end{table}
%%%%%%%%%%%%%%%%%%%%%%%%%%%%%%%%%%%%%%%%%%%%%%%%

An  arbitrarily polarized plane wave was considered to impinge normally on the face $z=0$ of the chosen metasurface. Hence, the incident electric field phasor can be written as
\begin{equation}
\#E_{inc}=\tond{E_{0x} \ux+E_{0y} \uy} \exp\tond{-i \ko z},\quad
z > 0.
\label{eq:Einc}
\end{equation}
For all calculations, we set either
\begin{itemize}
\item
$E_{0x}=1$~V~m$^{-1}$ and $E_{0y}=0$ for $x$-polarized incidence
or
\item
$E_{0x}=0$ and $E_{0y}=1$~V~m$^{-1}$   for $y$-polarized incidence.
\end{itemize}

The dimensions of the unit cell were chosen 
so that all nonspecular components of the transmitted field are evanescent.
Therefore, far from the face $z=-t-b$ as $z\to-\infty$, the transmitted electric
field phasor can be written as
\begin{eqnarray}
\nonumber
&&\#E_{tr}=\quadr{\tond{\tau_{xx}E_{0x}+\tau_{xy}E_{0y}} \ux+\tond{\tau_{yy}E_{0y}+\tau_{yx}E_{0x}} \uy}
\\
&&\qquad \times \exp\tond{ -i \ko z}, \quad z \to -\infty,
\label{eq:Etrasm}
\end{eqnarray}
where $\tau_{xx}$ and $\tau_{yy}$ are the co-polarized specular transmission coefficients whereas $\tau_{xy}$ and $\tau_{yx}$ are the cross-polarized specular transmission coefficients.

The  response characteristics of the chosen metasurface to the
normally incident plane wave were obtained using
the 3D electromagnetic simulator ANSYS\textregistered~\cite{Ansys}. 
The simulator considered only the unit cell with periodicity
 imposed with respect to both $x$ and $y$, Floquet theory \cite{Maystre,PMLbook}
being invoked to represent the $x$- and $y$-variations of the electric and magnetic field
phasors. 

The actual structure simulated is constituted by the unit cell  and two $a\times{a}\times{d}$ vacuous regions, one above and the other below
the unit cell along the $z$ axis, with $d\simeq\lambdao /4$.
The following four-step iterative procedure involving an adaptive mesh was used to obtain convergent results:
\begin{itemize}
\item[(i)] A mesh is generated with tetrahedrons of certain dimensions.
\item[(ii)]  The electromagnetic boundary-value problem is solved to determine
the scattering matrix of the analyzed structure. 
\item[(iii)]  Another mesh with smaller tetrahedrons is generated and step (ii)
is repeated.
\item[(iv)] A test is performed on the non-zero elements of the scattering matrixes
obtained in steps (ii) and (iii). If no element  changes by more than $0.1\%$ in magnitude,
the procedure is stopped. If not,  the second mesh is designated as the first mesh  a new iteration and steps (ii)--(iv) are repeated.
\end{itemize}
The iterative procedure was started with an initial
mesh comprising tetrahedrons having a 
maximum dimension  $0.2\lambdao$ and the tetrahedron dimensions were reduced by $20\%$ every iteration.  Convergence was reached within 12 iterations.

\section{Results and Discussions}\label{sec:RaD}
For all data reported here, we fixed  $a=18.36$~\SImum, 
 $b=2.5$~\SImum, $t=0.5$~\SImum,
 $w=3.6$~\SImum, $l=9.6$~\SImum,  $h=16.8$~\SImum, 
and $\eps_d=2.1$. Calculations were made for
$f=\omega/2\pi\in\quadr{0.5,5.5}$~THz with $T\in\graff{233,373}$~K
and $\Bo\in\graff{0,1}$~T. For all three configurations of
the magnetostatic field,   $\tau_{xy}$ and $\tau_{yx}$ 
turned out to be negligibly small. Hence, we present spectrums
of only $\vert\tau_{xx}\vert$ and $\vert\tau_{yy}\vert$ in this section.

\subsection{Stopbands of $\tau_{xx}$}
\subsubsection{First stopband}\label{3.A.1}
\begin{figure}[htbp]
	\begin{center}
		\begin{tabular}{c}
			(a) \includegraphics[width=75mm]{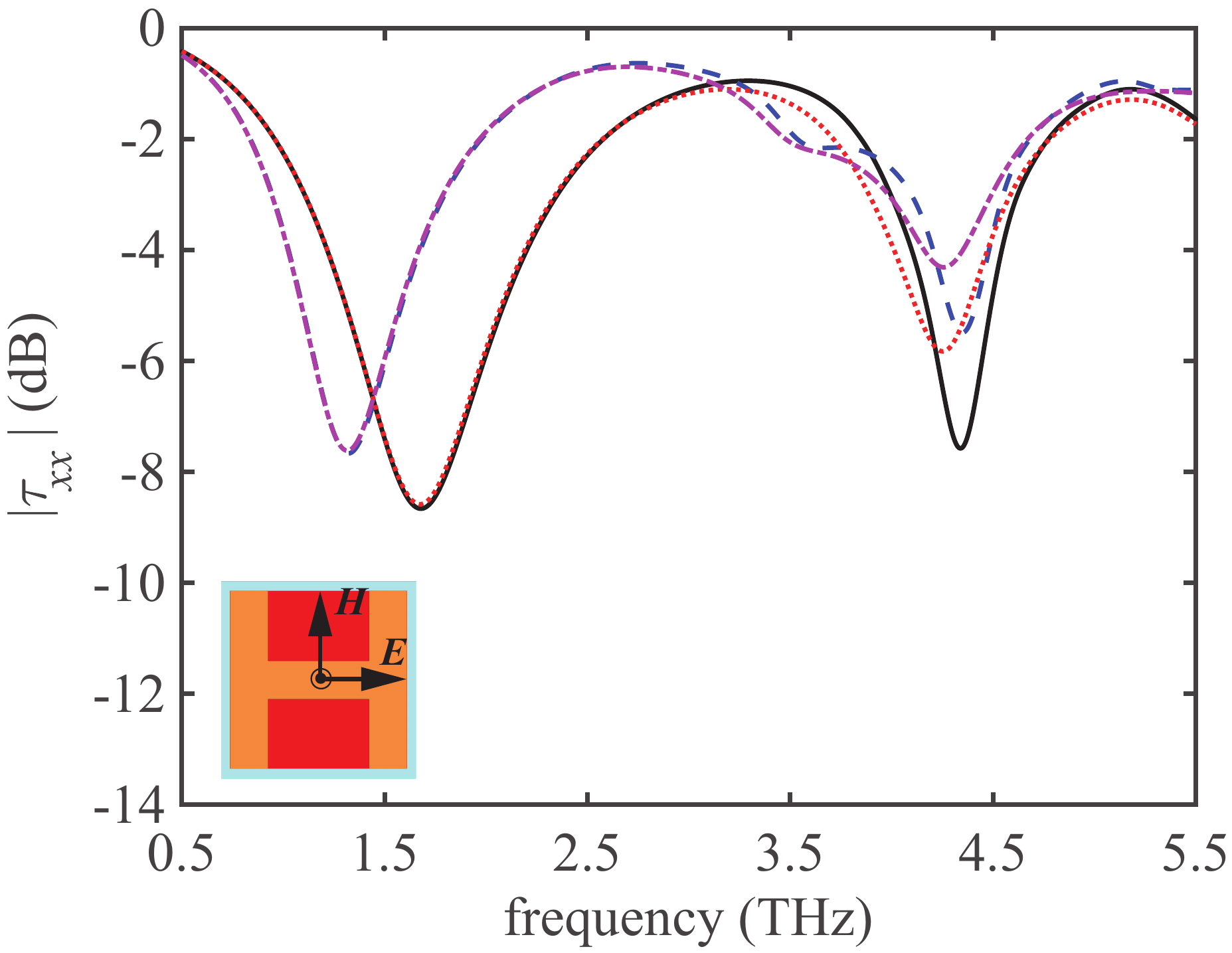}\\%\hspace{10pt}
			(b) \includegraphics[width=75mm]{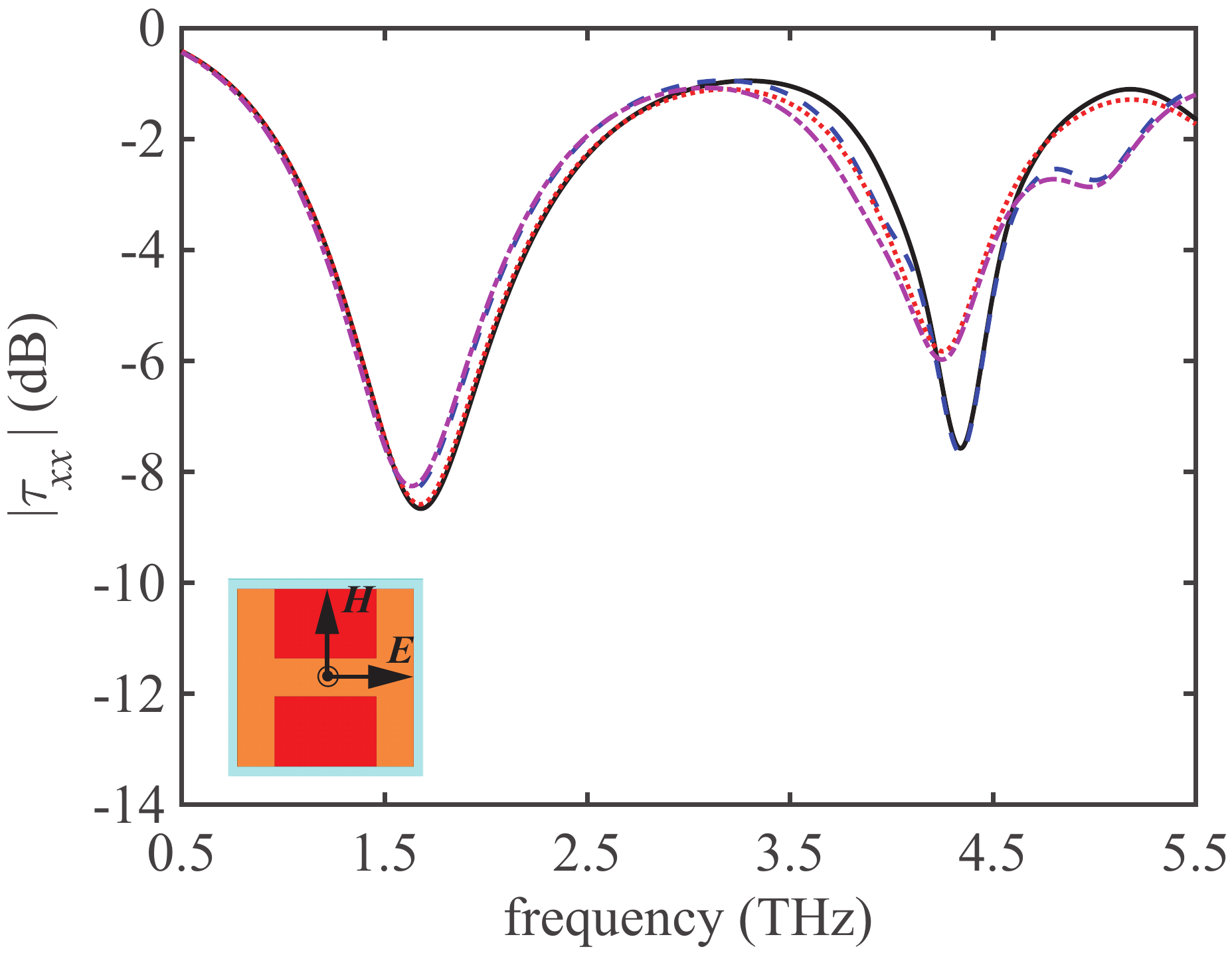}\\
			(c) \includegraphics[width=75mm]{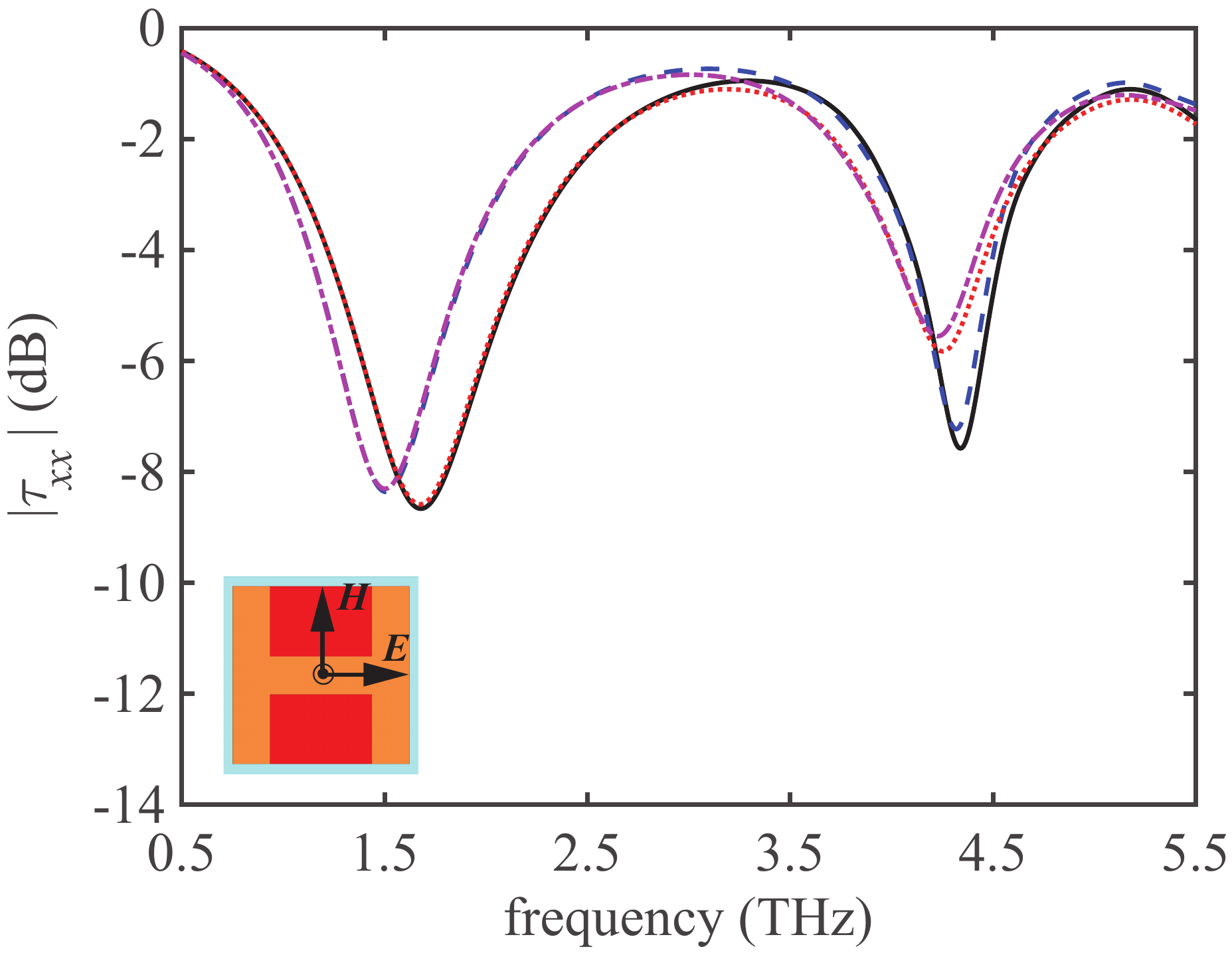}
		\end{tabular}
	\end{center}
	\caption{(Color online) Spectrums of $\vert\tau_{xx}\vert$ for (a) Faraday, (b) Voigt-X, and (c) Voigt-Y configurations.
	The solid black curves are for $B_0=0$~T and $T=233$~K, the dotted red curves
	for $B_0=0$~T and $T=373$~K, the dashed blue curves for $B_0=1$~T and $T=233$~K,
	and the dashed-dotted magenta curves for $B_0=1$~T and $T=373$~K. The insets  show the polarization state of the incident plane wave.}
	\label{fig:tau_xx}
\end{figure}

The spectrums of $\mod{\tau_{xx}}$  for all three distinctive configurations of the magnetostatic field are presented in
Fig.~\ref{fig:tau_xx} for  
 $\Bo\in\graff{0,1}$~T and $T\in\graff{233,373}$~K. 
 A prominent stopband (transmission less than $-7$~dB) exists between
 1.3 and 1.7~THz in all spectrums. Let the baseline environmental parameters
 be specified as $\Bo=0$ and $T=233$~K (black curves in
Fig.~\ref{fig:tau_xx}), for which the center frequency
 $\fo$ of the stopband is  $1.678$~THz. Obviously, since $\Bo=0$, this value
 of $\fo$  is the same for  all 
three magnetostatic-field configurations: Faraday [Fig.~\ref{fig:tau_xx}(a)], Voigt-X [Fig.~\ref{fig:tau_xx}(b)], and Voigt-Y [Fig.~\ref{fig:tau_xx}(c)]. Increasing the temperature to the high value ($T=373$~K)
results in a small redshift of the spectrums (red curves), with $\fo$ decreasing to  $1.674$~THz. The percentage relative shift $\Delta\fo/\fo=-0.24\%$ of the center frequency ($\Delta \fo=-4$~GHz)  must be due to the CdTe pixels because the effect of temperature  on the relative permittivity of InAs was assumed to be small enough to be ignored, and magnetostatic effects on the relative permittivity of CdTe were ignored similarly.

%%%%%%%%%%%%%% FIGURE 2 %%%%%%%%%%%%%%%%%%%%%%%%
%%%%%%%%%%%%%%%%%%%%%%%%%%%%%%%%%%%%%%%%%%%%%%%%

Increasing the magnitude of the magnetostatic field from $0$ to $1$~T instead of increasing the temperature produces a larger redshift (blue curves) with respect to the baseline (black curves). In the Faraday configuration, $\fo$ redshifts from   $1.678$~THz to $1.321$~THz when $\Bo$
changes from $0$ to $1$~T at a fixed temperature of $233$~K, the
 shift   $\Delta \fo=-357$~GHz, and a  percentage relative shift of $\Delta\fo/\fo=-21.28\%$, of the center frequency being huge.  The redshift of $\fo$ is less strong in the Voigt-Y configuration: from  $1.678$~THz to $1.507$~THz resulting in   $\Delta\fo/\fo=-10.19\%$. The redshift is even weaker for the Voigt-X configuration:
from  $1.678$~THz to $1.637$~THz resulting in $\Delta\fo/\fo=-2.44\%$. All three of these  shifts  must be due to the InAs pixels because the CdTe pixels cannot be affected by the magnetostatic field.

Switching on the $1$-T magnetostatic field   and increasing the temperature 
from $233$~K to $373$~K simultaneously invokes the magnetostatic controllability of the InAs pixels and the thermal controllability of the CdTe pixels in a cooperative fashion.  The center frequency $\fo$ of the  stopband shifts by $-362$~GHz, $-45$~GHz, and $-177$~GHz to
 $1.316$~THz,   $1.633$~THz, and $ 1.501$~THz, corresponding to a 
 percentage relative shift  $\Delta\fo/\fo=-21.57\%$, $-2.68\%$, and $-10.55\%$ for the Faraday, Voigt-X, and Voigt-Y configuration, respectively.
 
The central frequencies of the  stopband of $\tau_{xx}$  in the range 1.3--1.7-THz range for all magnetostatic-field configurations and chosen values of the magnetostatic field and temperature are provided in  Table~\ref{tab:tau_xx}. The data  indicate  (i) that the magnetostatic field gives coarse control whereas temperature gives fine control,
 and (ii) that   $\Bo$ and $T$ act cooperatively.

%%%%%%%%%%%%%% TABLE 2 %%%%%%%%%%%%%%%%%%%%%%%%
\begin{table*}[htbp]
	\caption{\bf Center frequencies (THz) of the first stopband (between
	 1.3 and 1.7~THz) of $\mod{\tau_{xx}}$ for $B_0=\graff{0,1}$~T and $T=\graff{233,373}$~K for the Faraday, Voigt-X, and Voigt-Y configurations.	\label{tab:tau_xx}
	}
	\begin{center} 
		\begin{tabular}{l|c|c|c|c}
			Magnetostatic & $\Bo=0$ T& $\Bo=0$ T & $\Bo=1$ T & $\Bo=1$ T \\
			configuration & $T=233$~K & $T=373$~K &$T=233$~K & $T=373$~K  \\
			\hline
			Faraday & 1.678 & 1.674 & 1.321 & 1.316\\
			Voigt-X & 1.678 & 1.674 & 1.637 & 1.633\\
			Voigt-Y & 1.678 & 1.674 & 1.507 & 1.501\\
			\hline%\hline
		\end{tabular}
	\end{center}
\end{table*}
%%%%%%%%%%%%%%%%%%%%%%%%%%%%%%%%%%%%%%%%%%%%%%%%

The metasurface response depends on the magnetostatic-field configuration because of the asymmetry of the $\sf{H}$ shape. Thus, rotation about the $z$ axis by 90 deg changes the resonator shape, allowing a distinction between the responses for the Voigt-X and Voigt-Y configurations. Rotation about
	the $z$ axis, however, can not affect the response for the Faraday configuration.
    
\subsubsection{Second stopband}\label{3.A.2}
The spectrums of $\mod{\tau_{xx}}$ presented in Fig.~\ref{fig:tau_xx} exhibit a second stopband in the  
4.2--4.4-THz range. 
When $\Bo=0$ and $T=233$~K, the center frequency
$\fo$ of the second stopband is  $4.340$~THz, as identified in Table~\ref{tab:tau_xx_2}.
Raising the temperature to $373$~K without turning on the magnetostatic field results in a shift of $-87$~GHz, the percentage relative bandwidth shift  $\Delta\fo/\fo=-2.00\%$ being about 22 times larger than of the first stopband. Thus, the thermal-control modality due to the CdTe pixels is more effective in the higher-frequency part of the $[0.5,5.5]$-THz range.

%%%%%%%%%%%%%% TABLE 3 %%%%%%%%%%%%%%%%%%%%%%%%
\begin{table*}[htbp]
	\caption{\bf Center frequencies (THz) of the second stopband  (between
	 4.2 and 4.4~THz) of $\mod{\tau_{xx}}$ for $B_0=\graff{0,1}$~T and $T=\graff{233,373}$~K for the Faraday, Voigt-X, and Voigt-Y configurations.	\label{tab:tau_xx_2}
	}
	\begin{center} 
		\begin{tabular}{l|c|c|c|c}
			Magnetostatic & $\Bo=0$ T& $\Bo=0$ T & $\Bo=1$ T & $\Bo=1$ T \\
			configuration & $T=233$~K & $T=373$~K &$T=233$~K & $T=373$~K  \\
			\hline
			Faraday & 4.340 & 4.253 & 4.349 & 4.258\\
			Voigt-X & 4.340 & 4.253 & 4.340 & 4.250\\
			Voigt-Y & 4.340 & 4.253 & 4.320 & 4.229\\
			\hline%\hline
		\end{tabular}
	\end{center}
\end{table*}
%%%%%%%%%%%%%%%%%%%%%%%%%%%%%%%%%%%%%%%%%%%%%%%%

The effect of the InAs pixels by themselves is different on the second stopband from that on the first stopband. In the Faraday configuration, $\fo$ blueshifts from   $4.340$~THz to $4.349$~THz when $\Bo$ changes from $0$ to $1$~T at a fixed temperature of $233$~K, so that the the percentage relative percentage  shift  $\Delta\fo/\fo=0.21\%$. Concurrently, no shift at all is evident for the Voigt-X configuration whereas
the shift is $ -20$~GHz (i.e., $\Delta\fo/\fo= {0.46}\%$) for the  Voigt-Y configuration.  In contrast, the first stopband redshifted by much larger margins for all three magnetostatic-field configurations.

By switching on the $1$-T magnetostatic field   and increasing the temperature 
from $233$~K to $373$~K simultaneously, the center frequency 
$\fo$ of the second stopband shifts by $-82$~GHz, $-90$~GHz, and $-11$~GHz to
 $4.258$~THz,  $ 4.250$~THz, and   $4.229$~THz corresponding to a 
 percentage relative shift $\Delta\fo/\fo=-1.89\%$, $-2.07\%$, and $-2.56\%$  for the Faraday, Voigt-X, and Voigt-Y configuration, respectively. All three shifts are redshifts, just the same
 as for the first stopband. However, the data in  Table~\ref{tab:tau_xx_2}  indicate  (i) that the magnetostatic field gives fine control whereas temperature gives coarse control,
 and (ii) that   $\Bo$ and $T$ act cooperatively for the Faraday and the Voigt-X configurations but not for the Voigt-Y configuration.

%   %%%%%%%%%%%%% TABLE 4 %%%%%%%%%%%%%%%%%%%%%%%%
%   \begin{table*}[htbp]
%   	\caption{\bf Spatial profiles of the electric and magnetic fields on the top surface of the unit cell
%	at the center frequency $\fo$ of the first stopband of $\tau_{xx}$, when the magnetostatic
%	field is in the Faraday configuration. The normalization factors are $E_{inc}=+\sqrt{\vert{E_{0x}}\vert^2+\vert{E_{0y}}\vert^2}$ and $H_{inc}=E_{inc}/\etao$.
%	\label{tab:field_xx}}
%   	\begin{center} 
%   		\begin{tabular}{c|c|c|c|c|c}
%   			& $\Bo=0$ T, $T=233$~K & $\Bo=0$ T, $T=373$~K & $\Bo=1$ T, $T=233$~K & $\Bo=1$ T, $T=373$~K \\
%   			& $\nu=1.678$~THz & $\nu=1.674$~THz & $\nu=1.321$ & $\nu=1.316$~THz
%   			\\
%   			\hline
%   			\hline
%   			\rule{0pt}{10\normalbaselineskip} 
%   \rotatebox{90}{\hspace{11mm}$\mod{\bf E}/E_{inc}$}
%   			&\includegraphics[width=0.2\linewidth]{E1678_F}
%   			 &\includegraphics[width=0.2\linewidth]{E1674_F}&  \includegraphics[width=0.2\linewidth]{E1321_F} & \includegraphics[width=0.2\linewidth]{E1316_F}& \includegraphics[height=36mm]{Escale} \\
%   			 \hline
%			\rule{0pt}{10\normalbaselineskip} 
%			\rotatebox{90}{\hspace{11mm}$\mod{{\bf H}}/H_{inc}$}
%			&\includegraphics[width=0.2\linewidth]{H1678_F}
%   			&\includegraphics[width=0.2\linewidth]{H1674_F}&  \includegraphics[width=0.2\linewidth]{H1321_F} & \includegraphics[width=0.2\linewidth]{H1316_F} & \includegraphics[height=36mm]{Hscale} \\
%   			\hline
%   		\end{tabular}
%   	\end{center}
%   \end{table*}

   %%%%%%%%%%%%%% TABLE 4 %%%%%%%%%%%%%%%%%%%%%%%%
\begin{table*}[htbp]
	\caption{\bf Spatial profiles of the electric and magnetic fields on the top surface of the unit cell
		at the center frequency $\fo$ of the first stopband of $\tau_{xx}$, when the magnetostatic
		field is in the Faraday configuration. The normalization factors are $ {E_{inc}}=+\sqrt{\vert{E_{0x}}\vert^2+\vert{E_{0y}}\vert^2}$ and $ {H_{inc}=E_{inc}}/\etao$.
		\label{tab:field_xx}}
	\begin{center} 
		\begin{tabular}{c}
\includegraphics[width=1.0\linewidth]{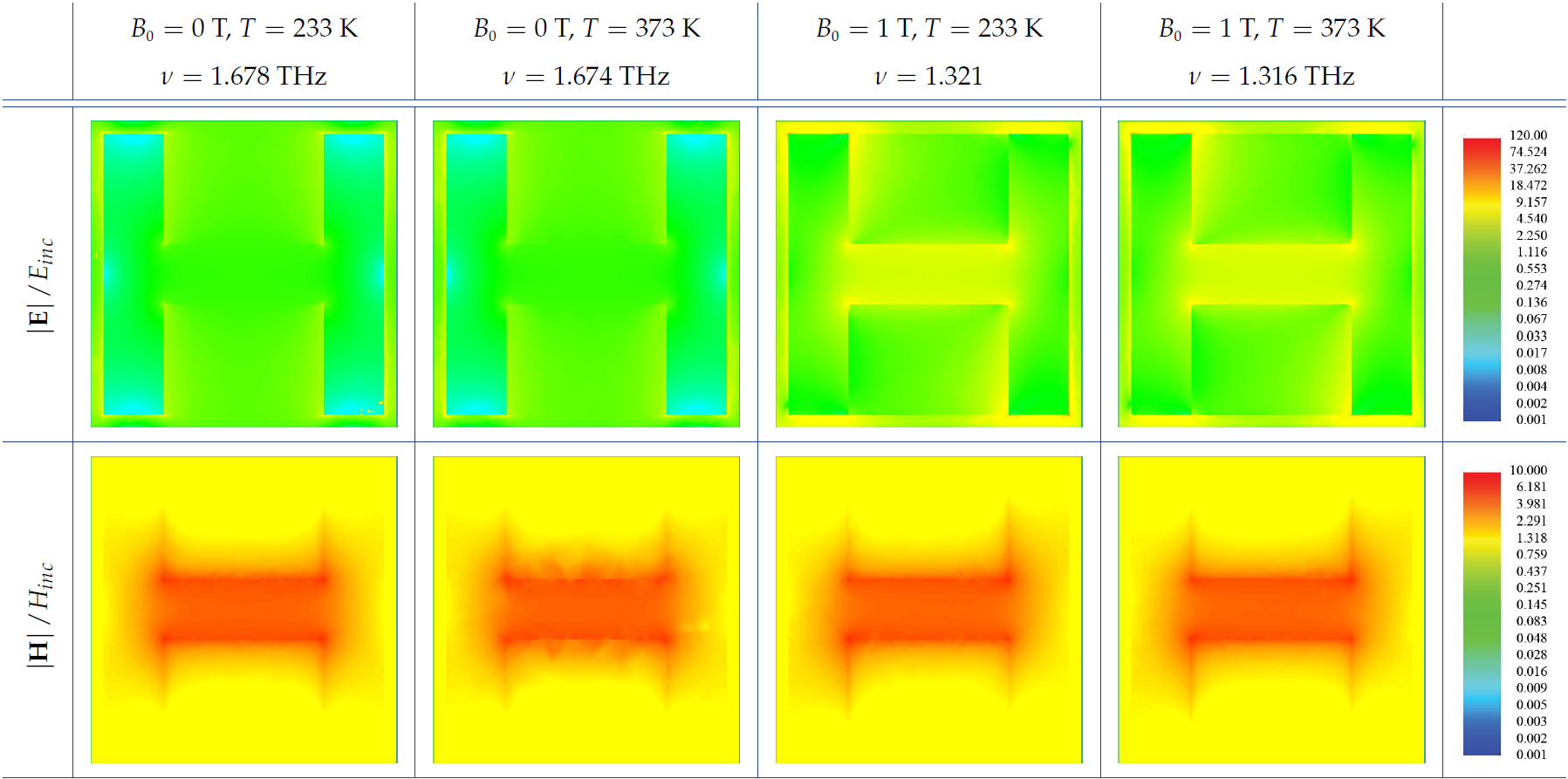}
		\end{tabular}
	\end{center}
\end{table*}

In order to  examine how the switching   the magnetostatic field on/off as well as the raising/lowering  the temperature affects the transmission spectrums, we calculated the field distributions at the center frequencies of the first stopbands of $\mod{\tau_{xx}}$ when the magnetostatic field is in the Faraday configuration. 
Spatial profiles of the electric
and magnetic fields at the center frequencies of the first stopband
are shown in Table~\ref{tab:field_xx}  for combinations of $B_0\in\left\{0,1\right\}$~T
and $T\in\left\{233,373\right\}$~K.
Clearly, a change in temperature affects the spatial profile of the magnetic field significantly, particularly in the CdTe region close to the central section of the 
$\sf H$ (made of InAs), but not much the spatial profile of the electric field. Likewise, a change in the magnetostatic field affects the spatial profile of the electric field significantly, particularly in the central section as well as at the extremities of both legs of the
$\sf H$, but  the spatial profile of the electric field is insignificantly affected.
In passing, let us note that at corners and edges, the electric field can be more than hundred times larger than the amplitude of the incident electric  field and the magnetic field
can be more ten times larger than the amplitude of the incident magnetic  field.

\subsection{Stopbands of $\tau_{yy}$}
\subsubsection{First stopband}\label{3.B.1}
\begin{figure}
	\begin{center}
		\begin{tabular}{c}
			(a) \includegraphics[width=75mm]{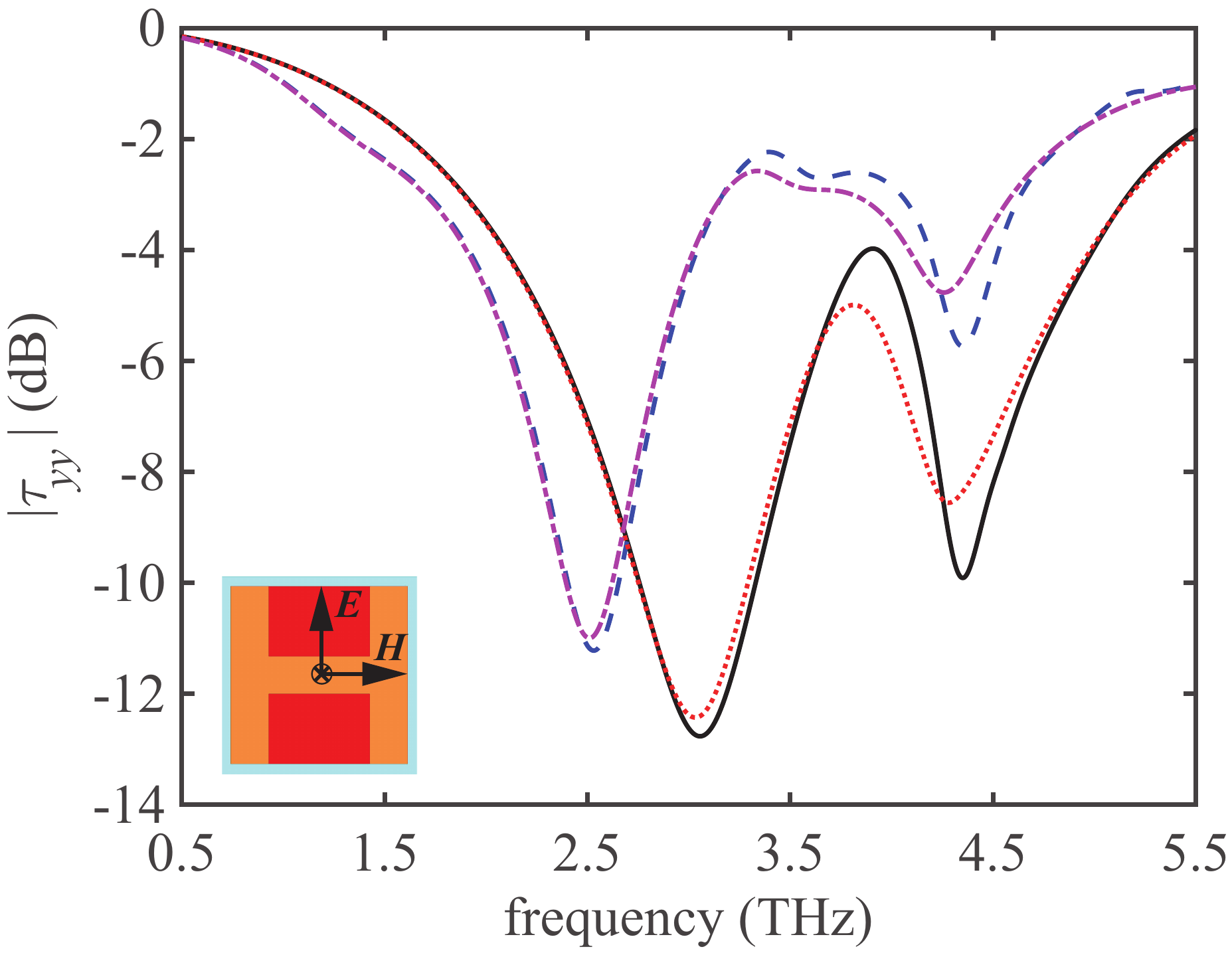}\\%\hspace{10pt}
			(b) \includegraphics[width=75mm]{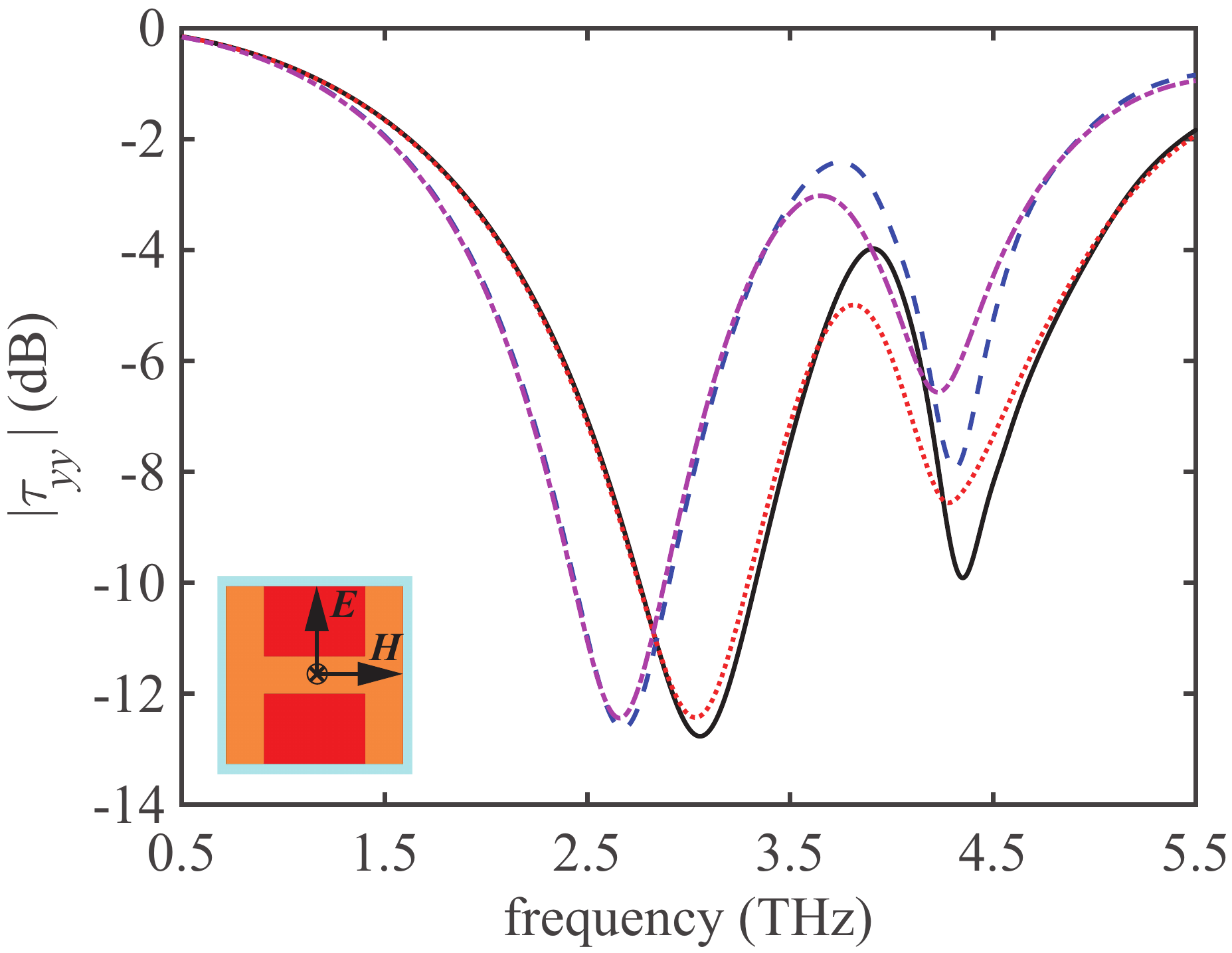}\\
			(c) \includegraphics[width=75mm]{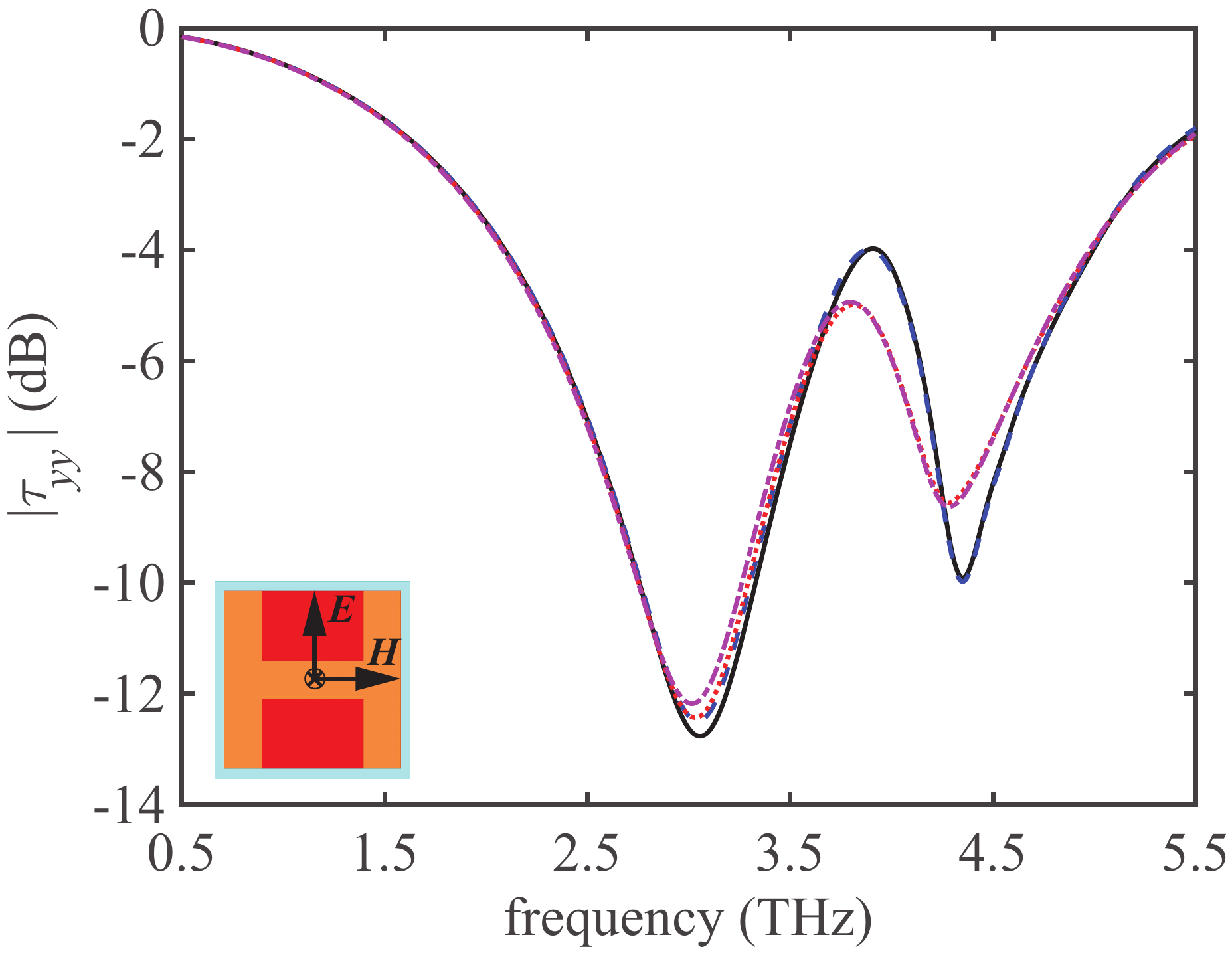}
		\end{tabular}
	\end{center}
	\caption{(Color online) Same as Fig.~\ref{fig:tau_xx}, but the spectrums
	of $\vert\tau_{yy}\vert$ are shown.
		}
	\label{fig:tau_yy}
\end{figure}

Figure~\ref{fig:tau_yy} is analogous to Fig.~\ref{fig:tau_xx}, except that the spectrums of $\mod{\tau_{yy}}$ instead of $\mod{\tau_{xx}}$  are plotted.
Similarly to what has been observed for $\mod{\tau_{xx}}$, $\mod{\tau_{yy}}$   exhibits a prominent stopband  (transmission less than $-11$~dB) between
 $2.5$ and $3.1$~THz. For the baseline condition
specified as $\Bo=0$ and $T=233$~K (black curves in
Fig.~\ref{fig:tau_yy}),   the center frequency
 $\fo$ of the stopband is  $3.055$~THz for all three magnetostatic-field configurations: Faraday [Fig.~\ref{fig:tau_yy}(a)], Voigt-X [Fig.~\ref{fig:tau_yy}(b)], and Voigt-Y [Fig.~\ref{fig:tau_yy}(c)]. Increasing the temperature from
 $233$~K to $373$~K
results in the redshift of $\fo$  from  $3.055$ to $3.032$~THz  due to the CdTe pixels. The  percentage relative shift being $\Delta\fo/\fo=-0.75\%$, the shift  $\Delta \fo=-23$~GHz  is about six times larger in magnitude than the one for $\tau_{xx}$.

%%%%%%%%%%%%%% FIGURE 3 %%%%%%%%%%%%%%%%%%%%%%%%
%%%%%%%%%%%%%%%%%%%%%%%%%%%%%%%%%%%%%%%%%%%%%%%%

Unlike what was found for $\tau_{xx}$, switching on the $1$-T magnetostatic field while keeping the temperature fixed at $233$~K produces a significantly larger redshift (blue curves) 
for $\tau_{yy}$ with respect to the baseline   for the Faraday and Voigt-X configurations but
 a significantly smaller redshift for the Voigt-Y configuration, as becomes
 clear from comparing Tables~\ref{tab:tau_xx} and \ref{tab:tau_yy}.
The shift is  $-524$~GHz, $-381$, and $-11$~GHz, corresponding to the percentage relative shift $\Delta\fo/\fo=-17.15\%$, $-12.47\%$, and $-0.36\%$, respectively, for
the Faraday, Voigt-X, and Voigt-Y configuration. These  shifts  are due to the InAs pixels.

  %%%%%%%%%%%%%% TABLE 5 %%%%%%%%%%%%%%%%%%%%%%%%
\begin{table*}[htbp]
	\caption{\bf  Center frequencies (THz) of the first stopband (between
		 $2.5$ and $3.1$~THz) of $\mod{\tau_{yy}}$ for $B_0=\graff{0,1}$~T and $T=\graff{233,373}$~K for the Faraday, Voigt-X, and Voigt-Y configurations.	\label{tab:tau_yy}
	}
	\begin{center} 
		\begin{tabular}{l|c|c|c|c}
	Magnetostatic& $\Bo=0$ T& $\Bo=0$ T & $\Bo=1$ T & $\Bo=1$ T \\
		configuration& $T=233$~K & $T=373$~K &$T=233$~K & $T=373$~K  \\
			\hline
Faraday & 3.055 &  3.032 &  2.531 &  2.510\\
Voigt-X & 3.055 &  3.032 &  2.674 &  2.659\\
Voigt-Y & 3.055 &  3.032 &  3.044 &  3.015\\
          \hline%\hline
		\end{tabular}
	\end{center}
\end{table*}
%%%%%%%%%%%%%%%%%%%%%%%%%%%%%%%%%%%%%%%%%%%%%%%%

By switching on the $1$-T magnetostatic field    and increasing the temperature 
from $233$~K to $373$~K simultaneously, even larger redshifts  become evident in the spectrums of $\mod{\tau_{yy}}$ (magenta curves) in Fig.~\ref{fig:tau_yy}.
The  central frequency $\fo$ of the stopband shifts to $2.510$~THz, $2.659$~THz, and  $3.015$~THz,  for the Faraday, Voigt-X, and Voigt-Y configuration, respectively, the shift being  $-545$~GHz, $-396$~GHz, and $-40$~GHz and the percentage relative  bandwidth shift is $\Delta\fo/\fo=-17.84\%$, $-12.96\%$, and $-1.31\%$, correspondingly.  Thus, both $\Bo$ and $T$ act cooperatively  for $\tau_{yy}$, just as for  $\tau_{xx}$.

\subsubsection{Second stopband}\label{3.B.2}
The spectrums of $\mod{\tau_{yy}}$ presented in Fig.~\ref{fig:tau_yy} exhibit a second stopband in the $4.2$--$4.4$~THz frequency range. For all three distinctive configurations of the magnetostatic field, unlike what  was observed for the first stopband in Sec.~3.\ref{3.B.1},  fine control comes from the magnetostatic field and  coarse control from the temperature, as may also be deduced from the data presented in Table~\ref{tab:tau_yy_2}.

%%%%%%%%%%%%%% TABLE 6 %%%%%%%%%%%%%%%%%%%%%%%%
\begin{table*}[htbp]
	\caption{\bf  Center frequencies (THz) of the second stopband (between
		 $4.2$ and $4.4$~THz) of $\mod{\tau_{yy}}$ for $B_0=\graff{0,1}$~T and $T=\graff{233,373}$~K for the Faraday, Voigt-X, and Voigt-Y configurations.	\label{tab:tau_yy_2}
	}
	\begin{center} 
		\begin{tabular}{l|c|c|c|c}
	Magnetostatic& $\Bo=0$ T& $\Bo=0$ T & $\Bo=1$ T & $\Bo=1$ T \\
		configuration& $T=233$~K & $T=373$~K &$T=233$~K & $T=373$~K  \\
			\hline
Faraday & 4.352 &  4.282 &  4.352 &  4.259\\
Voigt-X & 4.352 &  4.282 &  4.312 &  4.229\\
Voigt-Y & 4.352 &  4.282 &  4.351 &  4.285\\
          \hline%\hline
		\end{tabular}
	\end{center}
\end{table*}
%%%%%%%%%%%%%%%%%%%%%%%%%%%%%%%%%%%%%%%%%%%%%%%%

%%%%%%%%%%%%%% TABLE 7 %%%%%%%%%%%%%%%%%%%%%%%%
%\begin{table*}[htbp]
%	\caption{\bf Spatial profiles of the electric and magnetic fields on the top surface of the unit cell
%	at the center frequency $\fo$ of the first stopband of $\tau_{yy}$, when the magnetostatic
%	field is in the Voigt-X configuration. The normalization factors are $E_{inc}=+\sqrt{\vert{E_{0x}}\vert^2+\vert{E_{0y}}\vert^2}$ and $H_{inc}=E_{inc}/\etao$.
%\label{tab:field_yy}}
%	\begin{center} 
%		\begin{tabular}{c|c|c|c|c|c}
%			& $\Bo=0$ T, $T=233$~K & $\Bo=0$ T, $T=373$~K & $\Bo=1$ T, $T=233$~K & $\Bo=1$ T, $T=373$~K \\
%			& $\nu=3.055$~THz & $\nu=3.032$~THz & $\nu=2.674$ & $\nu=2.659$~THz\\
%			\hline
%			\hline
%			\rotatebox{90}{\hspace{11mm}$\mod{\bf E}/E_{inc}$}
%			&\includegraphics[width=0.2\linewidth]{E3055_VX}
%			&\includegraphics[width=0.2\linewidth]{E3032_VX}&  \includegraphics[width=0.2\linewidth]{E2674_VX} & \includegraphics[width=0.2\linewidth]{E2659_VX}& \includegraphics[height=36mm]{Escale} \\
%			\hline
%			\rotatebox{90}{\hspace{11mm}$\mod{\bf H}/H_{inc}$}
%			&\includegraphics[width=0.2\linewidth]{H3055_VX}
%			&\includegraphics[width=0.2\linewidth]{H3032_VX}&  \includegraphics[width=0.2\linewidth]{H2674_VX} & \includegraphics[width=0.2\linewidth]{H2659_VX} & \includegraphics[height=36mm]{Hscale} \\
%			\hline
%		\end{tabular}
%	\end{center}
%\end{table*}

%%%%%%%%%%%%% TABLE 7 %%%%%%%%%%%%%%%%%%%%%%%%
\begin{table*}[htbp]
	\caption{\bf Spatial profiles of the electric and magnetic fields on the top surface of the unit cell
		at the center frequency $\fo$ of the first stopband of $\tau_{yy}$, when the magnetostatic
		field is in the Voigt-X configuration. The normalization factors are $ {E_{inc}}=+\sqrt{\vert{E_{0x}}\vert^2+\vert{E_{0y}}\vert^2}$ and $ {H_{inc}=E_{inc}}/\etao$.
		\label{tab:field_yy}}
	\begin{center} 
		\begin{tabular}{c}
\includegraphics[width=1.0\linewidth]{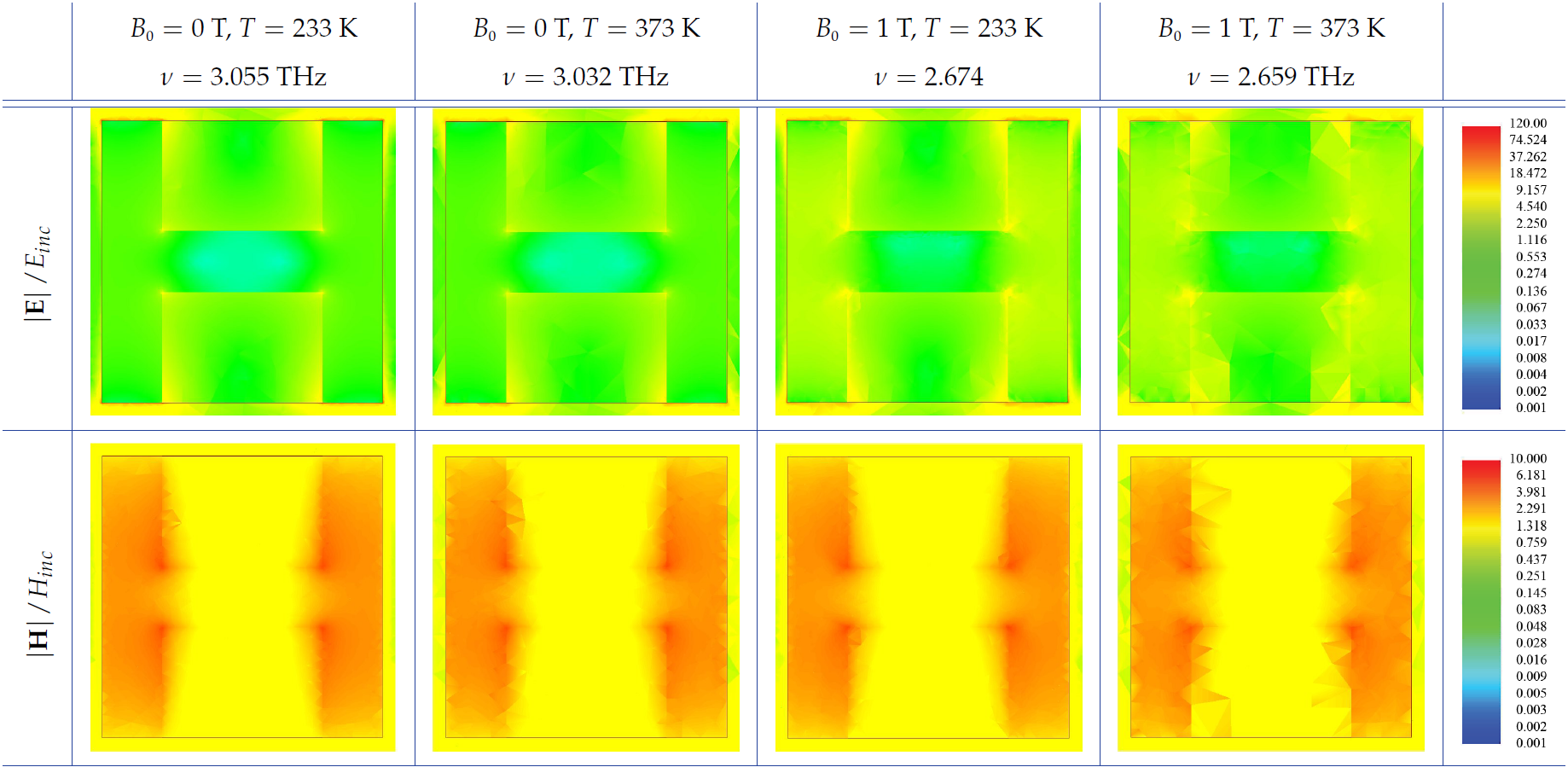}
		\end{tabular}
	\end{center}
\end{table*}

Indeed, for the baseline condition specified as $\Bo=0$ and $T=233$~K (black curves in
Fig.~\ref{fig:tau_yy}),   the center frequency $\fo$ of the stopband is  $4.352$~THz for all three magnetostatic-field configurations. Increasing the temperature from $233$~K to $373$~K while keeping the magnetostatic field  null valued
results in the redshift of $\fo$  from  $4.352$ to $4.282$~THz, i.e.,    $\Delta\fo/\fo=-1.61\%$ and $\Delta \fo=-70$~GHz. This shift is comparable to the shift observed for the second stopband in
the spectrum of $\tau_{xx}$ in Sec.~3.\ref{3.A.2}.
Switching on the $1$-T magnetostatic field instead of increasing the temperature from $233$~K produces a shift of $0$~GHz, $-40$, and $-1$ GHz, with a percentage relative   shift  $\Delta\fo/\fo=0\%$, $-0.92\%$, and $-0.02\%$, correspondingly. respectively, for the Faraday, Voigt-X, and Voigt-Y configuration. Clearly, the thermal control modality is much more effective than the magnetostatic control modality.

Switching on the $1$-T magnetostatic field   and increasing the temperature 
from $233$~K to $373$~K simultaneously makes the center frequency of the  second stopband shift by $-93$~GHz,  $-123$, and $-67$~GHz to $4.259$~THz,  
 $4.229$~THz, and $4.285$~THz giving a percentage relative  shift $\Delta\fo/\fo=-2.14\%$, $-2.83\%$, and $-1.54\%$. for the Faraday, Voigt-X, and Voigt-Y configuration, respectively. 
We conclude that     $\Bo$ and $T$ act cooperatively.

Spatial profiles of the electric
and magnetic fields at the center frequencies of the first stopband
are shown in Table~\ref{tab:field_yy}  for combinations of $B_0\in\left\{0,1\right\}$~T
and $T\in\left\{233,373\right\}$~K for the Voigt-X configurations.
Clearly, a change in the magnetostatic field affects the spatial profile of the magnetic field significantly, particularly in the central section and both legs of the
$\sf H$ made of InAs, but the spatial profile of the electric field is not affected
much. Likewise, a change in temperature field affects the spatial profile of the electric field significantly, particularly in the CdTe regions close to the legs of the
$\sf H$, but not significantly the spatial profile of the magnetic field.

\section{Concluding remarks}\label{sec:cr}
We have theoretically substantiated the concept of multicontrollability for metasurfaces by employing two differently controllable materials in the $\sf{H}$-shaped
subwavelength scattering elements of a specific metasurface. The transmission spectrums of the chosen metasurface exhibit prominent stopbands in the THz regime. These stopbands shift when either thermally controllable pixels in the scattering elements are influenced by increasing the temperature or the magnetostatically controllable pixels  in the scattering elements are influenced by turning on a magnetostatic field.    Depending on the spectral location of
the stopband, either
the magnetostatic field gives coarse control and temperature gives fine 
control or \textit{vice versa}. The level of magnetostatic control depends on the
magnetostatic-field configuration. The largest   shifts emerge when both control modalities are simultaneously deployed. 

Other control modalities---such as electrical, optical,  piezoelectric, and magnetostrictive \cite{Gersten}---can be invoked by using pixels made of diverse materials~\cite{AkhSPIE}. Numerous geometries are possible for the subwavelength scattering elements \cite{Walia,Kivshar}. We plan to report our further work on multicontrollable metasurfaces in appropriate forums.

\vspace{5mm}

\noindent {\bf Acknowledgment.}
  AL thanks the Charles Godfrey Binder Endowment at Penn State for ongoing support of his research.

\end{document}